\documentclass[a4paper,11pt]{article}
\usepackage{amsfonts}
\usepackage[T1]{fontenc}
\usepackage{graphicx}
\usepackage[colorlinks]{hyperref}
\usepackage{amsmath}
\usepackage{amssymb}
\usepackage{latexsym}
\usepackage{color}
\usepackage{cite}
\usepackage{float}
\usepackage[textfont={footnotesize,sf},labelfont={color=blue,bf,sf},labelsep=endash]{caption}
\usepackage[position=top,labelfont={color=blue,bf,sf}]{subfig}
\usepackage{bm}
\usepackage{makeidx}
\usepackage{colortbl}
\usepackage{csquotes}
\usepackage[squaren]{SIunits}
\newcommand{\bra}{\begin{array}}
\newcommand{\era}{\end{array}}
\newcommand{\beq}{\begin{equation}}
\newcommand{\eeq}{\end{equation}}
\newcommand{\beqar}{\begin{eqnarray}}
\newcommand{\eeqar}{\end{eqnarray}}

\def\BC{\bb C}
\def\_\BC{\bbi C}

\def\Tr {{\rm Tr}}

\def\( {\left(}
   \def\) {\right)}
\def\[ {\left[}
\def\] {\right]}

\def\Tr {{\rm Tr}}

\newcommand{\lb}{\label}

\def\eqref#1{\textcolor{blue}{(\ref{#1})}}
\usepackage{amsmath}

\usepackage{xcolor}
\colorlet{linkequation}{blue}
\usepackage[colorlinks]{hyperref}

\newcommand*{\SavedEqref}{}
\let\SavedEqref\eqref
\renewcommand*{\eqref}[1]{%
  \begingroup
    \hypersetup{
      linkcolor=linkequation,
      linkbordercolor=linkequation,
    }%
    \SavedEqref{#1}%
  \endgroup
}
\hypersetup{linkcolor=black}
\begin{document}
\begin{titlepage}
\setcounter{page}{1}
\renewcommand{\thefootnote}{\fnsymbol{footnote}}


\vspace{5mm}
\begin{center}

{\Large \bf {Entanglement in a two-qubit Heisenberg XXX model
with x-components of Dzyaloshinskii-Moriya and Kaplan-Shekhtman-Entin-Wohlman-Aharony interactions }}

\vspace{4mm}

{\bf Rachid Hou\c{c}a$^{1,2}$\footnote{r.houca@uiz.ac.ma}
,
El Bouâzzaoui Choubabi$^{2}$\footnote{choubabi.e@ucd.ac.ma}
,
Mohammed El Bouziani$^{2}$\footnote{elbouziani.m@ucd.ac.ma}
,
Abdelhadi Belouad$^{2}$\footnote{belabdelhadi@gmail.com}
and
Abdellatif Kamal$^{2,3}$\footnote{abdellatif.kamal@ensam-casa.ma}
}

\vspace{3mm}

{$^{1}$\em Team of Theoretical Physics and High Energy, Department of Physics, Faculty of Sciences, Ibn
Zohr University}, Agadir, Morocco,\\
{\em PO Box 8106, Agadir, Morocco}

{$^{2}$\em Team of Theoretical Physics, Laboratory L.P.M.C., Department of Physics, Faculty of Sciences, Chouaib
Doukkali University, El Jadida, Morocco},\\
{\em PO Box 20, 24000 El Jadida, Morocco}

{$^{3}$\em ISPS2I Laboratory, National Higher School of Arts and Crafts (ENSAM), Hassan II University, 20670, Casablanca, Morocco}

\vspace{2.5cm}

\begin{abstract}
This article explores the concurrence in a two-qubit Heisenberg XXX model with Dzyaloshinskii-Moriya (DM) and Kaplan-Shekhtman-Entin-Wohlman-Aharony (KSEA) interactions. The concurrence expression was developed using the physical variables connected with the chosen system. Our results indicate that the temperature, the spin coupling constant, the x-components of the DM and KSEA interactions may all play a role in determining the degree of intricacy between states. Additionally, these findings imply that the separability of states is obtained at high-temperature domains or by switching the spin coupling. In contrast, the entanglement of states may be achieved at low temperatures or by using high values of the x-components of the DM and KSEA parameters. Furthermore, the DM and KSEA interactions have an identical effect on the concurrence behaviors at high temperatures.
\end{abstract}
\end{center}

\vspace{2cm}



\end{titlepage}

\section{Introduction}
Because of its nonlocality \cite{Schr,EPR,bell}, {\color{red}entanglements have} become an important tool in quantum information processing \cite{Bennett1,Murao}, including quantum computing \cite{Grover1,Grover2}, quantum cloning \cite{Chiara} and teleportation \cite{Yeo}. Using the Heisenberg model as a basic system, entangled states may be generated and controlled. This model has been used to simulate a wide range of physical phenomena, including quantum dots \cite{Trauzettel}, nuclear spins \cite{Kane}, optical lattices \cite{Sorensen}, and superconductors \cite{Nishiyama}. Entanglement in a Heisenberg system with spin is more important because of several magnetic properties of high-spin quantum systems, such as the Haldane gap \cite{Morra,Wang}. It has been shown that the Heisenberg model in the two-site system has a thermal entanglement for a given spin, such as the spin-$1/2$ chain \cite{Zhou}, spin-$1$ chain \cite{Song}, and $(1/2,3/2)$ mixed-spin chain \cite{Guo}.

Many quantum entanglement measurements have been established. The most important of which is that of Bennett \textit{et al.}, who developed quantitative measures of quantum entanglement in bipartite systems, first for pure states \cite{Ben1} and then for hybrid states \cite{Ben2}. These measures utilize entropy and do not use the extremization technique to get the optimum working ensemble. Because extremization is notoriously tricky to deal with theoretically, the authors could only obtain a precise formula for entanglement formation in Bell-diagonal states \cite{Woo}. Wootters then proved Hill and Wootters' theory, which offers a clear prescript for assessing the entanglement of any two-qubit system \cite{Hil}. The Wootters formula requires the solution of a fourth-degree algebraic problem, requiring tedious Ferrari formulas. Thus, concise formulations for multiple quantum states are required for practical reasons.

Researchers in quantum information have recently begun to investigate the Dzyaloshinskii-Moriya (DM) interaction, which was discovered by Dzyaloshinskii and Moriya and named after them. The scientific community has shown a significant interest in the DM interaction, which is thought to be one of the most effective regulating characteristics of quantum entanglement. Condensed matter systems with correlated components have recently received much interest. As a result, recent work has focused on quantifying and characterizing quantum correlations in Heisenberg spin-chain models with DM interaction \cite{JAF1,JAF2,JAF3}. The results show that the DM interaction may control quantum correlations.
The non-classical thermal correlations of the two-qubit Heisenberg chain subjected to a transverse magnetic field and DM interaction were also investigated using the local quantum uncertainty (LQU) \cite{22}. Pourkarimi investigated the influence of the DM interaction on quantum discord in the Heisenberg XY spin chain model \cite{Pourkarimi}. His findings indicate that quantum discord is more persistent than quantum entanglement in the presence of the DM interaction. G. L. Kamta \textit{et al.} explore the entanglement of a two-qubit anisotropic Heisenberg XY chain in thermal equilibrium at temperature $T$ in an external magnetic field $B$. They produce entanglement for any finite T through the combined influences of anisotropic interactions and a magnetic field B by adjusting its strength  \cite{Kamta}. Sun \textit{et al.} demonstrated that increasing the DM interaction value increases the quantum discord of the two-qubit Heisenberg XYZ spin chain model \cite{Sun}. N. Canosa \textit{et al.} examine the entanglement of general mixed states of a two-qubit Heisenberg XYZ chain in a magnetic field and its detection through different criteria. The exact separability and weaker conditions implied by the disorder and the von Neumann entropic criteria are analyzed \cite{Canosa}.

In the fundamental study, \cite{Moriya1,Moriya2}, it was discovered that there is symmetric helical coupling associated with the DM interaction. It is ignored since the symmetric helical interaction is significantly weaker than the DM interaction. Kaplan discovers this symmetric helical interaction between local spins in the single-band Hubbard model with spin-orbit couplings (SOC) \cite{Kaplan}. Following that, Shekhtman, Entin-Wohlman, and Aharony demonstrate that the mild ferromagnetism of La2CuO4 may be explained by this non-zero symmetric helical interaction \cite{Shekhtman}. As a result, the symmetric helical interaction was named the Kaplan-Shekhtman-Entin-Wohlman-Aharony (KSEA) interaction. Thermal quantum entanglement and discord have been investigated in a two-qubit XYZ Heisenberg model with DM, and KSEA interactions in an inhomogeneous magnetic field \cite{Yur}. The two-qubit Heisenberg XYZ model with the interactions of DM and KSEA in thermal equilibrium was studied by A. V. Fedorova \textit{et al.}  \cite{fedrova}. They derived analytical formulas for the LQU, and local quantum Fisher information (LQFI) uses available expressions for the entropic quantum discord. By using the correlation functions and chiral order factors, H. Fu \textit{et al.} investigated the impact of the KSEA interaction on the ground state features of three kinds of spin chains in a transverse field by using the correlation functions and chiral order factors \cite{fu}.

Low-dimensional magnetism physics has recently focused on finding novel magnetic compounds and refining their properties to fulfil emerging technology demands. Knowing that integrable models of magnetism in low-dimensional materials are based
on the homogeneous Heisenberg XXX model and its anisotropic version XXZ \cite{Motamedifar}.  Motivated by the cited works and the utility of Heisenberg XXX models, we investigated the thermal entanglement in a two-qubit homogeneous Heisenberg XXX model with x-component of DM and KSEA interactions.

The organization of this paper is as follows. In section \textcolor[rgb]{0.00,0.00,1.00}{2}, in detail, we present the theoretical model of our system subjected to DM and KSEA interactions. In addition, we determine the ground states and their entanglement at absolute zero temperature. Section \textcolor[rgb]{0.00,0.00,1.00}{3} will be devoted to the thermal quantum entanglement and concurrence expression with some limiting cases. In section \textcolor[rgb]{0.00,0.00,1.00}{4}, numerical studies will be performed to highlight the system behavior. Finally, the paper is closed with a findings overview and our perspectives on the studied system.

\section{Theoretical model}
We consider a two-qubits Heisenberg XXX model with DM and KSEA interactions of a one-half isotropic Heisenberg XXX model. The  Hamiltonian of the system can be written as:
\beq
\mathcal{H}=\mathcal{H}_H+\mathcal{H}_{DM}+\mathcal{H}_{KSEA}
\eeq
where the first term refers to Heisenberg exchange couplings, the second one denotes the DM interaction, and the third one represents the KSEA interaction. Generally, the DM Hamiltonian $\mathcal{H}_{DM}$ \cite{Moriya1,Moriya2} can be expressed as follows
\begin{eqnarray}
\mathcal{H}_{DM} &=& \boldsymbol{D}. \left(\boldsymbol{\sigma}_1\times\boldsymbol{\sigma}_{2}\right) \\ \nonumber
 &=& D_x\left(\sigma_1^y\sigma_{2}^z-\sigma_1^z\sigma_{2}^y\right)+D_y\left(\sigma_1^x\sigma_{2}^z-\sigma_1^z\sigma_{2}^x\right)+
 D_z\left(\sigma_1^x\sigma_{2}^y-\sigma_1^y\sigma_{2}^x\right)
\end{eqnarray}
where $\boldsymbol{D} = (D_x, D_y, D_z)$ reflects the DM interaction vector and $\sigma^{x,y,z}$ represents the Pauli spin matrices. However, the KSEA Hamiltonian \cite{Kaplan,Shekhtman} may be expressed as
\begin{eqnarray}
\mathcal{H}_{KSEA} &=& \boldsymbol{\sigma}_1.\boldsymbol{\Gamma}.\boldsymbol{\sigma}_{2} \\ \nonumber
                   &=& (\sigma_1^x, \sigma_1^y, \sigma_1^z) \left(
                                            \begin{array}{ccc}
                                              0 & \Gamma_z & \Gamma_y \\
                                              \Gamma_z & 0 & \Gamma_x \\
                                              \Gamma_y & \Gamma_x & 0 \\
                                            \end{array}
                                          \right)\left(
                                                   \begin{array}{c}
                                                     \sigma_2^x \\
                                                     \sigma_2^y \\
                                                     \sigma_2^z \\
                                                   \end{array}
                                                 \right)\\ \nonumber
                  &=& \Gamma_x\left(\sigma_1^y\sigma_{2}^z+\sigma_1^z\sigma_{2}^y\right)+\Gamma_y\left(\sigma_1^x\sigma_{2}^z+\sigma_1^z\sigma_{2}^x\right)+
 \Gamma_z\left(\sigma_1^x\sigma_{2}^y+\sigma_1^y\sigma_{2}^x\right)
\end{eqnarray}
where $\boldsymbol{\Gamma}$ denotes the traceless symmetric tensor. Thus, both the DM and KSEA interactions are generally composed of three components. {\color{red} In the present work, we thought to investigate the effect of x-component interactions on thermal entanglements. To do this, we restrict ourselves to the case where the DM and KSEA interactions occur along the x-axis.} Then, our Hamiltonian can be reads as
\beq\lb{fr}
\mathcal{H}=J \left(\sigma_1^x\sigma_{2}^x+\sigma_1^y\sigma_{2}^y+\sigma_1^z\sigma_{2}^z\right)+D_x \left(\sigma_n^y\sigma_{n+1}^z-\sigma_n^z\sigma_{n+1}^y\right)+\Gamma_x  \left(\sigma_n^y\sigma_{n+1}^z+\sigma_n^z\sigma_{n+1}^y\right)
\eeq
where $J$ represents the real coupling constant for the spin interaction. Antiferromagnetic chains are when the value of $J > 0$, while ferromagnetic chains are those in which the value of $J <0$. It should be noted that $D_x$ signifies the x-component of the DM interaction, while $\Gamma_x$ denotes the x-component of the KSEA interaction. The Hamiltonian  \eqref{fr} can be expressed in the standard computing basis  $|00>$, $|01 >$, $|10 >$, $|11 >$ as
\beq\lb{1}
\mathcal{H}=\left(
\begin{array}{cccc}
 J & - i \Gamma_x +i D_x & - i \Gamma_x -i D_x & 0 \\
  i \Gamma_x -i D_x & -J & 2 J & i \Gamma_x +i D_x \\
  i \Gamma_x +i D_x & 2 J & -J & i \Gamma_x -i D_x \\
 0 &- i \Gamma_x -i D_x & - i \Gamma_x +i D_x & J \\
\end{array}
\right)
\eeq
The eigenvalue equation's solution yields to the eigenvalues
\begin{eqnarray}
\epsilon_{1,2} &=&J\pm2 \Gamma_x \lb{22}\\
\epsilon_{3,4} &=& -J\pm 2 \eta\lb{222}
\end{eqnarray}
where $\eta=\sqrt{D_x^2+J^2}$, and the related eigenvectors
\begin{eqnarray}\lb{psi}
  |\varphi_1\rangle &=&  \frac{1}{2}|00\rangle -\frac{i}{2}|01\rangle-\frac{i}{2}|10\rangle+\frac{1}{2}|11\rangle \\ \nonumber
  |\varphi_2\rangle &=& \frac{1}{2}|00\rangle+\frac{i}{2}|01\rangle+\frac{i}{2}|10\rangle+\frac{1}{2}|11\rangle \\ \nonumber
  |\varphi_3\rangle &=& -\frac{1}{\sqrt{2}}\sin (\theta_1)|00\rangle-\frac{i}{\sqrt{2}}\cos (\theta_1)|01\rangle+\frac{i}{\sqrt{2}}\cos (\theta_1)|10\rangle+ \frac{1}{\sqrt{2}}\sin (\theta_1)|11\rangle \\ \nonumber
  |\varphi_4\rangle &=&-\frac{1}{\sqrt{2}}\sin (\theta_2)|00\rangle+\frac{i}{\sqrt{2}}\cos (\theta_2)|01\rangle-\frac{i}{\sqrt{2}}\cos (\theta_2)|10\rangle+ \frac{1}{\sqrt{2}}\sin (\theta_2)|11\rangle
\end{eqnarray}
where $\theta_{1,2}$ are defined by 
\begin{eqnarray}
  \theta_{1,2} &=&\arctan\left(\frac{D_x}{\eta\mp J}\right)
\end{eqnarray}
First, we address the entanglement of the system's ground state at absolute zero temperature, which is necessary before we can explore thermal quantum {\color{red}entanglements}. It is reasonable to suppose that the ground state entanglement occurs in both the antiferromagnetic and ferromagnetic instances since the energies are the exchange coupling $J$ functions.

In order to highlight the dependency of the ground state energy on KSEA coupling constant $\Gamma_x$, we explore three significant cases. The first one is $\Gamma_x>0$. Using equations \eqref{22} and \eqref{222}, the ground state energies can be written as
\begin{eqnarray}
\epsilon_{2} &=&J-2 \Gamma_x,  \qquad \text{if} \quad  \Gamma_x>J+\eta\\ \nonumber
\epsilon_{4} &=& -J- 2 \eta, \qquad \text{if} \quad \Gamma_x<J+\eta
\end{eqnarray}
Consequently, for $\Gamma_x>J+\eta$, the ground state it will be the entangled state $|\varphi_2\rangle$, regardless of the sign of $J$, and when $\Gamma_x<J+\eta$, the ground state is the entangled state $|\varphi_4\rangle$ independently of the sign of $J$. Because $|\varphi_2\rangle$ and $|\varphi_4\rangle$ are maximally entangled at low temperatures, thus conducting to the maximum concurrence of $\mathcal{C}=1$ is achieved.

In the second case, independently of the sign of $J$ for $\Gamma_x<0$, the ground state energies  are given by
\begin{eqnarray}
\epsilon_{1} &=&J+2 \Gamma_x,  \qquad \text{if} \quad  \Gamma_x<-J-\eta\\ \nonumber
\epsilon_{4} &=& -J- 2 \eta, \qquad \text{if} \quad \Gamma_x>-J-\eta
\end{eqnarray}
where $|\varphi_1\rangle$ and $|\varphi_4\rangle$ are  the corresponding maximally entangled states leading to the maximum concurrence $\mathcal{C}=1$.

\begin{figure}[!h]
  \centering
  \includegraphics[width=16 cm]{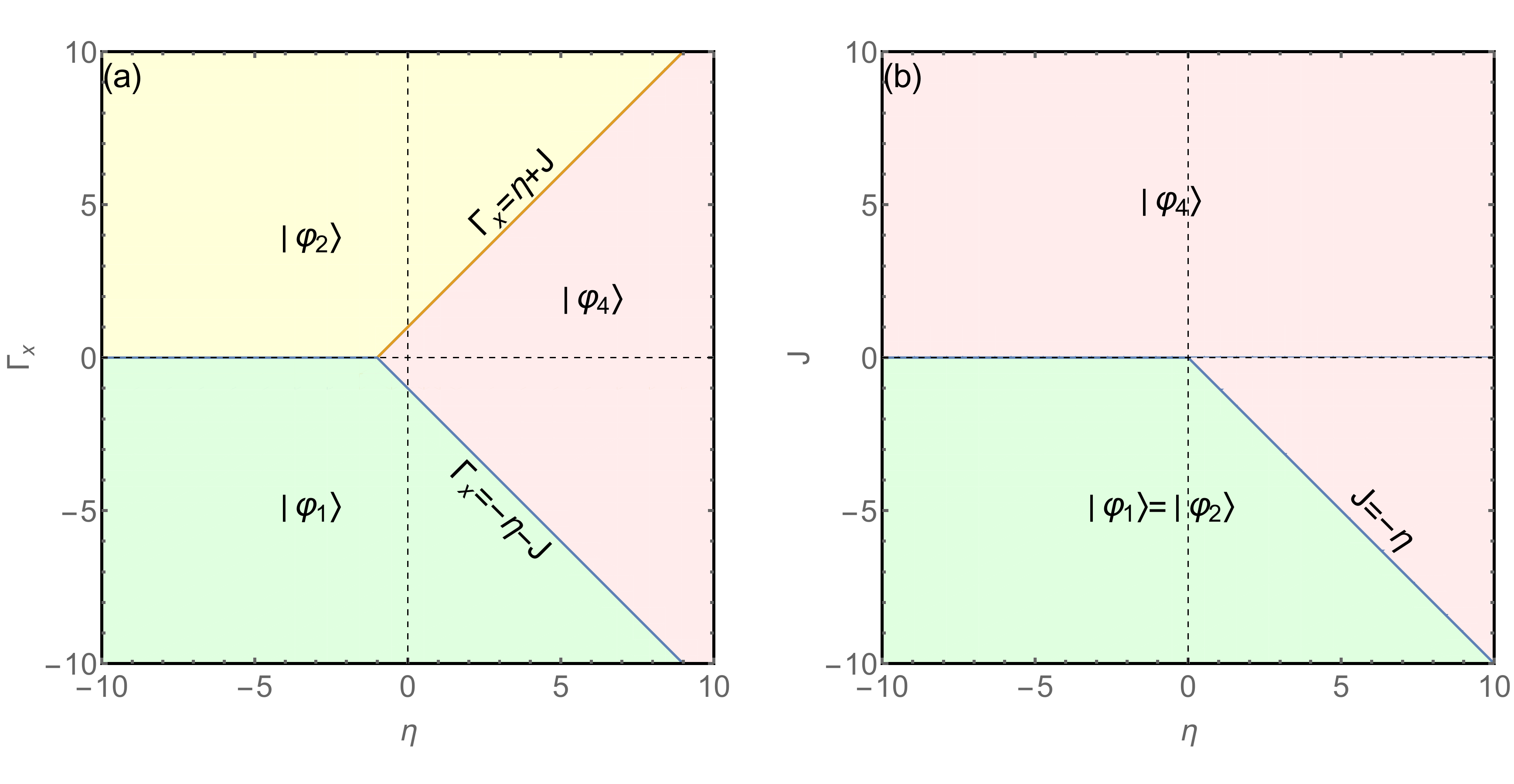}
  \caption{(Color online) Phase diagram at $T=0$ of a two qubits XXX Heisenberg
model. (a) $\Gamma_x\neq0$ for $J=1$,   (b) $\Gamma_x=0$.}\label{diag}
\end{figure}

However, in the last case where $\Gamma_x=0$, the ground state energies are dependent on the sign of $J$. In addition, if $J<0$, those energies are given by
\begin{eqnarray}
\epsilon_{1} &=& \epsilon_{2}=J,  \qquad \text{if} \quad  J<-\eta\\ \nonumber
\epsilon_{4} &=& -J- 2 \eta, \qquad \text{if} \quad J>-\eta.
\end{eqnarray}
In contrast,  for $J>0$, the ground state energy is given by
\begin{eqnarray}
\epsilon_{4} &=& -J- 2\eta, \qquad \text{for any value of } \eta
\end{eqnarray}
For the antiferromagnetic chains, the significant ground state is $|\varphi 4\rangle$. It was also the ground state for ferromagnetic chains when $J$ fulfilled the condition $J+\eta>0$, leading to maximal concurrence $\mathcal{C}=1$. To complete this investigation, we illustrate the three cases discussed above in the phase diagram shown in Fig. \ref{diag}.

\section{Thermal quantum entanglements}
After determining our system's spectrum, using the density matrix $\rho(T)$ to depict the system's state on thermal equilibrium at a given temperature $T$. The expression of $\rho(T)$ is presented by
\beq
\rho(T)={1\over\mathbb{Z}}e^{-\beta \mathcal{H}}
\eeq
where
\beq
\mathbb{Z}=\Tr e^{-\beta \mathcal{H}}
\eeq
Where $\mathbb{Z}$ is the canonical ensemble partition function and $\beta=1/ k_BT$ is the inverse thermodynamic temperature, where $k_B$ is the Boltzmann's constant, which is treated as unity in the following for simplicity. That will be succeeded by using the spectral decomposition of the Hamiltonian \eqref{1}, which allows the thermal density matrix $\rho(T)$ to be represented as
\beq\lb{3}
\rho(T)={1\over\mathbb{Z}}\sum_{l=1}^{4}e^{-\beta \epsilon_l}|\phi_l\rangle\langle\phi_l|
\eeq
The density matrix of the system, as discussed before in thermal equilibrium, may be represented in the normal computational basis by inserting \eqref{22} and \eqref{psi} into equation \eqref{3} and obtaining
\beq\lb{43}
\rho(T)={1\over\mathbb{Z}}\left(
\begin{array}{cccc}
 a & i \mu & i \nu & c \\
 -i\mu & b & d & -i\nu \\
 -i\nu & d & b & -i\mu \\
 c & i\nu & i\mu & a \\
\end{array}
\right)
\eeq
the elements matrix is represented by the equations
\begin{eqnarray}\lb{23}
a &=&\frac{1}{4} \left(2 e^{-\beta  \epsilon _3} \sin ^2\left(\theta _1\right)+2 e^{-\beta  \epsilon _4} \sin ^2\left(\theta _2\right)+e^{-\beta  \epsilon _1}+e^{-\beta  \epsilon _2}\right) \\ \nonumber
b &=& \frac{1}{4} \left(2 e^{-\beta  \epsilon _3} \cos ^2\left(\theta _1\right)+2 e^{-\beta  \epsilon _4} \cos ^2\left(\theta _2\right)+e^{-\beta  \epsilon _1}+e^{-\beta  \epsilon _2}\right) \\ \nonumber
c &=& \frac{1}{4} \left(-2 e^{-\beta  \epsilon _3} \sin ^2\left(\theta _1\right)-2 e^{-\beta  \epsilon _4} \sin ^2\left(\theta _2\right)+e^{-\beta  \epsilon _1}+e^{-\beta  \epsilon _2}\right) \\ \nonumber
d &=&\frac{1}{4} \left(-2 e^{-\beta  \epsilon _3} \cos ^2\left(\theta _1\right)-2 e^{-\beta  \epsilon _4} \cos ^2\left(\theta _2\right)+e^{-\beta  \epsilon _1}+e^{-\beta  \epsilon _2}\right)\\ \nonumber
\mu &=& -\frac{1}{4} \left(-e^{-\beta  \epsilon _3} \sin \left(2 \theta _1\right)+e^{-\beta  \epsilon _4} \sin \left(2 \theta _2\right)+e^{-\beta  \epsilon _1}-e^{-\beta  \epsilon _2}\right) \\ \nonumber
\nu &=&-\frac{1}{4} \left(e^{-\beta  \epsilon _3} \sin \left(2 \theta _1\right)-e^{-\beta  \epsilon _4} \sin \left(2 \theta _2\right)+e^{-\beta  \epsilon _1}-e^{-\beta  \epsilon _2}\right)
\end{eqnarray}
Consequently, the partition function is clearly described by
\beq
\mathbb{Z}=2 e^{\beta  J} \cosh \left(2 \beta  \eta\right)+2 e^{-\beta  J} \cosh (2 \beta  \Gamma_x )
\eeq
This matrix $\rho$ is examined in order to find out how much entanglement is linked with it.
\beq\lb{RR}
R=\rho S\rho^\ast S
\eeq
where $\rho$ has a complex conjugate in the form of $\rho^{\ast}$, and $S$ is determined by the formula
\beq
S=\sigma^y\otimes\sigma^y
\eeq
where $\sigma^y$ denotes the Pauli matrix, and the $R$ matrix may be obtained by a simple calculation by
\beq\lb{rr}
R=\left(
\begin{array}{cccc}
 R_{11} & R_{12} & R_{13} & R_{14}  \\
 R_{12}^\ast & R_{22} & R_{23}  & R_{13}^\ast \\
 R_{13}^\ast  & R_{23}  & R_{22} & R_{12}^\ast \\
 R_{14}  & R_{13} & R_{12} & R_{11} \\
\end{array}
\right)
\eeq
where the $R$ components matrix are denoted by
\begin{eqnarray}\lb{rrr}
  R_{11} &=& a^2+c^2+\mu ^2+\nu ^2 \\ \nonumber
  R_{22} &=& b^2+d^2+\mu^2+\nu^2\\ \nonumber
  R_{14} &=& 2ac+2\mu\nu \\ \nonumber
  R_{12} &=& i(\mu(a+b)+\nu  (c+d)) \\ \nonumber
  R_{13} &=& i (\nu(a+b)+\mu(c+d)) \\ \nonumber
  R_{23} &=& 2bd+2\mu\nu
\end{eqnarray}
To grasp the meaning of $R$, consider that $\Tr R$, which ranges from $0$ to $1$, measures the degree of equality between $\rho$ and $\rho^\ast$, which indicates how closely $\rho$ approximates a mixture of generalized Bell states. Additionally, the eigenvalues of $R$ are invariant under local unitary transformations of the individual qubits, which qualifies them to be included in a formula for entanglement, as entanglement must also be invariant under such transformations. Using \eqref{rr} and \eqref{rrr}, one can check quickly that the square roots of the eigenvalues of the matrix $R$ are given by

\begin{eqnarray}\lb{R}
\lambda_{1} &=& \left(\frac{(a-c)^2+(b-d)^2+2 (\mu -\nu )^2-(a+b-c-d) \sqrt{(a-b-c+d)^2+4 (\mu -\nu )^2}}{2 \mathbb{Z}^2}\right)^{1\over2} \\ \nonumber
\lambda_{2} &=& \left(\frac{(a-c)^2+(b-d)^2+2 (\mu -\nu )^2+(a+b-c-d) \sqrt{(a-b-c+d)^2+4 (\mu -\nu )^2}}{2 \mathbb{Z}^2}\right)^{1\over2} \\ \nonumber
\lambda_{3} &=& \left(\frac{(a+c)^2+(b+d)^2+2 (\mu +\nu )^2-(a+b+c+d) \sqrt{(a-b+c-d)^2+4 (\mu +\nu )^2}}{2 \mathbb{Z}^2}\right)^{1\over2} \\ \nonumber
\lambda_{4} &=& \left(\frac{(a+c)^2+(b+d)^2+2 (\mu +\nu )^2+(a+b+c+d) \sqrt{(a-b+c-d)^2+4 (\mu +\nu )^2}}{2 \mathbb{Z}^2}\right)^{1\over2}
\end{eqnarray}

When attempting to quantify the amount of entanglement associated with $\rho$, we take into consideration the concurrence \cite{Woo,Hil}, which may be described as follows:
\beq
\mathcal{C}=\max\left[0,2\max\left(\lambda_1,\lambda_2,\lambda_3,\lambda_4\right)-\sum_{i=1}^4\lambda_{i}\right]
\eeq
The expression mentioned above is significant in determining the extent of entanglement between states; however, the concurrence values are included in the range from zero, which represents an unentangled state, to one, which signifies a maximally entangled state. After deriving the concurrence
expression, which is implicitly reliant on the coupling constant $J$, the x-components of the DM and KSEA interactions, and the temperature $T$. Consequently, we now have all of the components necessary to examine the behavior of the proposed system concerning the previously specified quantities.
Before beginning the numerical section, it is needed to explore a few limiting cases to simplify the expression of concurrence and understand the behavior of our system.

The first limiting case concerns the high-temperature regime. To make things easier, we will use the same formulas for the various expressions in the previous section, assuming that $J=\Gamma_x=D_x=1$. The eigenvalues of the matrix $R$ are represented in the following manner:
\begin{eqnarray}
  \lambda_1 &\simeq& \left(\frac{1}{8} \left(1-2 \sqrt{2}\right) \beta +\frac{1}{16}\right)^{1\over2} \\ \nonumber
  \lambda_2 &\simeq&\left(\frac{1}{8} \left(2 \sqrt{2}+1\right) \beta +\frac{1}{16}\right)^{1\over2} \\
  \lambda_3 &\simeq& \left(\frac{1}{16}-\frac{3 \beta }{8}\right)^{1\over2} \\ \nonumber
  \lambda_4 &\simeq&\left(\frac{\beta }{8}+\frac{1}{16}\right)^{1\over2} \nonumber
\end{eqnarray}
Following a comparison of the eigenvalues of $R$ at high temperature, the expression of concurrence, in this case, maybe expressed by the relationship
\beq\lb{HT}
\mathcal{C}\simeq\max \left(0,\frac{1}{2} \left(\left(2 \sqrt{2}+1\right) \beta -1\right)\right)
\eeq
When $T$ approaches a substantial value, the expression of concurrence reveals that $\mathcal{C}\rightarrow0$ makes the system states separable and not entangled at high temperatures.

Strong exchange coupling $J$ is the second limiting case where we consider that $J\gg\Gamma_x$ and $J\gg D_x$. As a result, the quantities $\theta_1$ and $\theta_2$ have the values $\theta_1\simeq\pi/2$ and $\theta_2\simeq D_x/2 J$, respectively, and the eigenvalues of the matrix $R$ are dependent on the signature of $J$.  For an antiferromagnetic system ($J>0$), the eigenvalues of the matrix $R$ are represented by
 \begin{eqnarray}
  \lambda_1 &\simeq& \frac{1}{\left(2 \cosh (2 \beta  \Gamma_x )+e^{4 \beta  J}+1\right)} \\ \nonumber
  \lambda_2 &\simeq& \frac{e^{4 \beta  J}}{\left(2 \cosh (2 \beta  \Gamma_x )+e^{4 \beta  J}+1\right)} \\ \nonumber
  \lambda_3 &\simeq& \frac{1}{\left(2 \cosh (2 \beta  \Gamma_x )+e^{4 \beta  J}+1\right)} \\ \nonumber
  \lambda_4 &\simeq& \frac{1}{\left(2 \cosh (2 \beta  \Gamma_x )+e^{4 \beta  J}+1\right)} \nonumber
\end{eqnarray}
 and consequently, the expression of concurrence is given by
\beq
\mathcal{C}\simeq\max \left(0,\frac{e^{4 \beta  J}-3}{\left(2 \cosh (2 \beta  \Gamma_x )+e^{4 \beta  J}+1\right)}\right)
\eeq
If we look at the previous concurrence, it is easily possible to see that when $J\rightarrow+\infty$ , the value of the concurrence is $\mathcal{C}\rightarrow1$, which indicates that the states are maximally entangled. For $J\rightarrow0$, it is clear that the concurrence tends to $0$, which suggests that the state's system has become more separable. For a ferromagnetic system ($J<0$), the expression of concurrence is expressed by
\beq
\mathcal{C}\simeq\max \left(0,-\frac{e^{4 \beta  J}+1}{\left(2 \cosh (2 \beta  \Gamma_x )+e^{4 \beta  J}+1\right)}\right)
\eeq
The previous concurrence can have a value of $\mathcal{C}\rightarrow1$ when $J\rightarrow-\infty$, indicating that the states are most entangled. When $J\rightarrow0$, it is evident that the concurrence goes to $0$, meaning that the state's system has become more separable.

Finally, the last limiting case is studying strong DM and KSEA interactions. In this perspective, we assume that $D_x\gg J $ and $\Gamma_x \gg J$, which implies that the quantities $\theta _1$ and $\theta _2$ take on the values of $\theta_1=\theta_2\simeq \pi/4$, and as a result, the eigenvalues of the matrix $R$ are expressed in the following form
\begin{eqnarray}\lb{dfd}
  \lambda_1 &\simeq& \frac{e^{- \beta  D_x}}{\left(2 \left( \cosh \left(2 \beta  D_x\right)+ \cosh \left(2 \beta  \Gamma _x\right)\right)\right)^{1\over2}}\\ \nonumber
  \lambda_2 &\simeq& \frac{e^{ \beta  D_x}}{\left(2 \left( \cosh \left(2 \beta  D_x\right)+ \cosh \left(2 \beta  \Gamma _x\right)\right)\right)^{1\over2}}\\ \nonumber
  \lambda_3 &\simeq& \frac{e^{- \beta  \Gamma _x}}{\left(2 \left( \cosh \left(2 \beta  D_x\right)+ \cosh \left(2 \beta  \Gamma _x\right)\right)\right)^{1\over2}}\\ \nonumber
  \lambda_4 &\simeq& \frac{e^{ \beta  \Gamma _x}}{\left(2 \left( \cosh \left(2 \beta  D_x\right)+ \cosh \left(2 \beta  \Gamma _x\right)\right)\right)^{1\over2}}
\end{eqnarray}
It is clear that the \eqref{dfd} depends on $\Gamma_x$ and $D_x$. Therefore, the expression of concurrence is also dependent on the  $\Gamma_x$ and $D_x$. For $\Gamma_x >D_x$, the concurrence is expressed by
\beq
\mathcal{C}\simeq\max \left(0,\frac{\sqrt{2}\left(\sinh \left(\beta  \Gamma _x\right)-\cosh \left(\beta  D_x\right)\right)}{\left(\cosh \left(2 \beta  D_x\right)+\cosh \left(2 \beta  \Gamma _x\right)\right)^{1\over2}}\right)
\eeq
The last equation shows that the concurrence tends toward  asymptotic value $\mathcal{C}\simeq 1$ for strong KSEA interaction parameters $\Gamma_x$, which means that the states are  maximally entangled in this case. For  $\Gamma_x\rightarrow0$, the concurrence tends to $0$, which indicates that the states are separated for this set of limits. In the second case, for $\Gamma_x <D_x$, the concurrence  takes the form
\beq
\mathcal{C}\simeq\max \left(0,\frac{\sqrt{2}\left(\sinh \left(2 \beta  D_x\right)-\cosh \left(2 \beta  \Gamma _x\right)\right)}{\left(\cosh \left(2 \beta  D_x\right)+\cosh \left(2 \beta  \Gamma _x\right)\right)^{1\over2}}\right)
\eeq
When $D_x\rightarrow+\infty$, the concurrence takes the value $\mathcal{C}\simeq 1$, which indicates that the states are entangled. Inversely, when $D_x\rightarrow0$, the concurrence approaches $0$, reflecting states' separability for this specific condition.

To illustrate the suggested model's overall performance, we will devote the following section to a numerical evaluation of the concurrence $\mathcal{C}$, described below. Following that, we supplied a few plots in terms of the parameters of the considered system, such as temperature, spin coupling exchange, DM, and KSEA interactions, and we will have further discussions to conclude this.
\section{Numerical results}
This section of the paper will quantitatively investigate several features of entanglement in a two-qubit Heisenberg XXX chain with x-components of DM and KSEA interactions. First, we will examine concurrence as a function of temperature $T$ and the constant coupling exchange for the spin interaction $J$ by setting the x-component of DM and KSEA interactions to the value $D_x=\Gamma_x=1$. Second, we will set the coupling constants for the spin coupling exchange and the x-component of the KSEA interaction to $\Gamma_x=J=1$ and plot the concurrence $\mathcal{C}$ as a function of the x-component of the DM interaction $D_x$. Finally, we investigate the concurrence's behavior regarding the x-component of the KSEA interaction with fixed values $J=D_x=1$. For simplicity, the values $k_B=\hbar=1$ shall be assumed.
\begin{figure}[!h]
  \centering{\includegraphics[width=17.5
  cm]{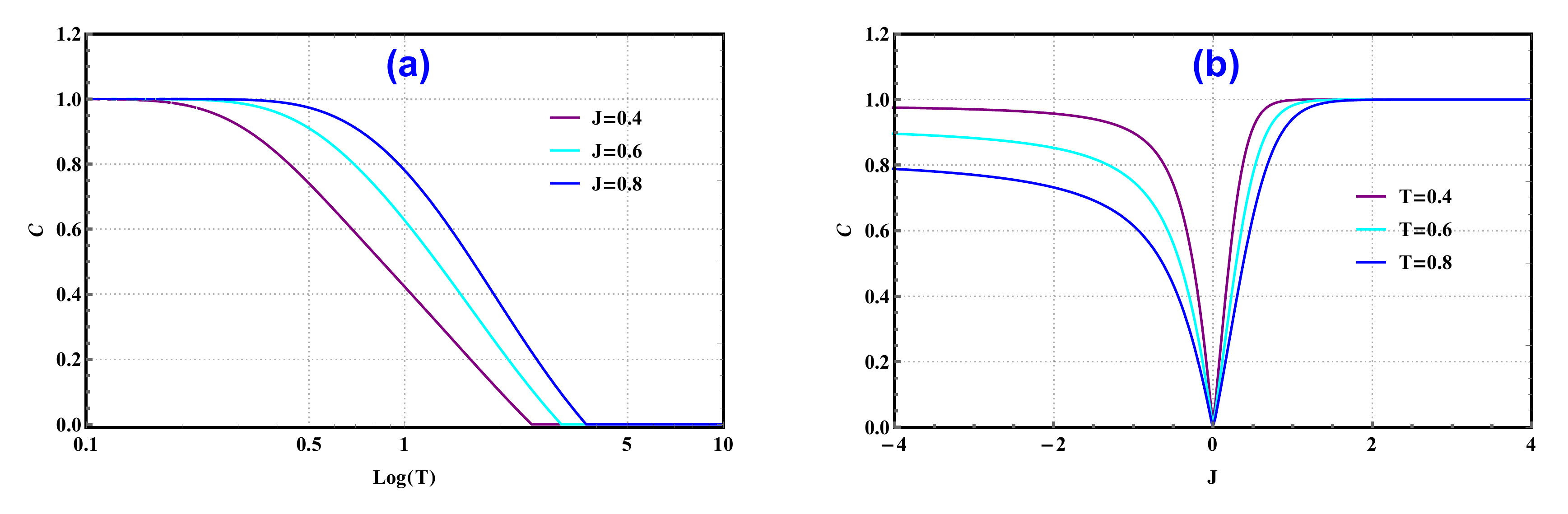}}
  \caption{(Color online) (a) The concurrence versus logarithmic temperature for different $J$. (b) The concurrence versus of constant coupling for different $T$ ($\Gamma_x=D_x=1$).}\label{f1}
\end{figure}

In Fig. \ref{f1}, we display the concurrence $\mathcal{C}$ in terms of logarithmic temperature $Log(T)$ for various coupling constants for the spin interaction $J=$ 0.4, 0.6 and 0.8 (Fig. \ref{f1}.a). And versus coupling constant for the spin interaction $J$ for different values of temperature $T=$  0.4, 0.6 and 0.8 (Fig. \ref{f1}.b). From Fig. \ref{f1}.a, one can observe threshold temperatures $T_c$ beyond which the entanglement disappears. The critical temperature $T_c$ is also impacted by the parameter $J$, such that $T_c$ and $J$ grow in proportion. Something more significant is that in the low-temperature and precisely in the interval $0<Log(T)<0.3$, although the value of $J$ rises, the entanglement stays equal to $1$, this means that in the vicinity of $ T = 0 $, the temperature is significant in front of the coupling. In addition, with the increase in temperature, we can achieve long-lasting entanglement by increasing the coupling between the spin (the blue curve).In Fig. \ref{f1}.b, for $J = 0$, we have a degeneracy for the ground state seen in the line delimiting separate ground states in Fig. 1.b. {\color{red}In this case, the ground state is a statistical mixture of the two states $|\varphi_2\rangle$ and $|\varphi_4\rangle$ which has zero concurrences.} For small values of $J$, the entanglement of the ferromagnetic and antiferromagnetic chains increases in the same way. According to Fig. \ref{f1}.b, we can see that the concurrence $\mathcal{C}$ for an antiferromagnetic system tends towards one rapidly for large values of $J$. In contrast, for a ferromagnetic chain and large values of $|J|$, the concurrence tends towards one, but, in comparison to antiferromagnetic systems, this progress is being made slowly. In conclusion, our system's states are more entangled at low temperatures or for large values of $|J|$; but, at high temperatures or for $ J =0 $, the system's states are more separable.
\begin{figure}[!h]
  \centering
  \includegraphics[width=17.5cm]{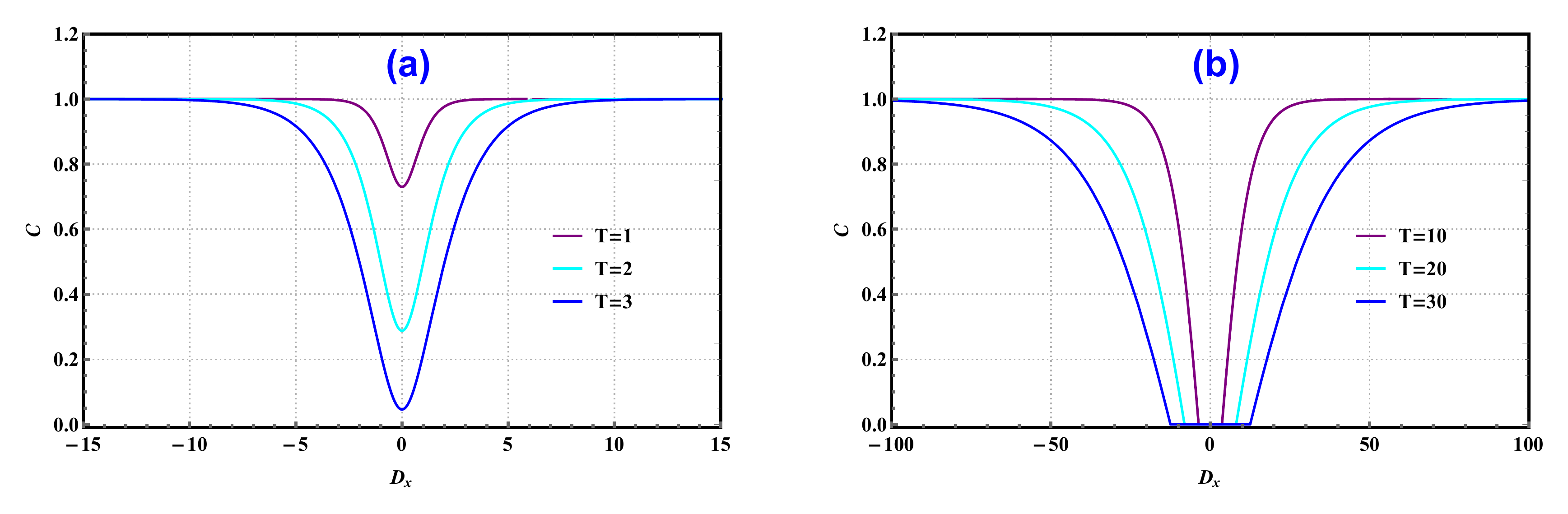}
  \caption{(Color online) The concurrence versus x-component of DM interaction for different $T$ ($J=1$ and $\Gamma_x=1$).}\label{f2}
\end{figure}

In Fig. \ref{f2}, we plot the concurrence $\mathcal{C}$ in terms of x-component $D_x$ at low temperature for $T=$ 1, 2 and 3 (a) and at high temperature for $T=$ 10, 20 and 30 (b). From Fig. \ref{f2}, the first observation is that the concurrence is symmetrical for $D_x=0$. The concurrence tends to a fixed value $\mathcal{C}=1$ for the high absolute value of the x-component of DM interaction even as the temperature rises. Moreover, for $D_x = 0$, we notice that the $\mathcal{C}$ presents a minimum at low temperature, which disappears with the decrease in temperature and vice versa, confirmed by the previous study in section (3). In addition, we note that the entanglement increases symmetrically, and with the decrease in temperature, we may get a short time entanglement. Otherwise, the concurrence shows an interval  T-dependent, where $ \mathcal{C} $ stays zero at high temperatures near $ D_x = 0 $. Hence, an increase in temperature indicates a widening of this interval. Then we can conclude that the state's system is influenced by the DM interaction, in which the large values of $D_x$ make the system more entangled. At high temperatures for small values of $D_x$, the system becomes less entangled, and the system's states become completely separable.

\begin{figure}[!h]
  \centering
  \includegraphics[width=17.5cm]{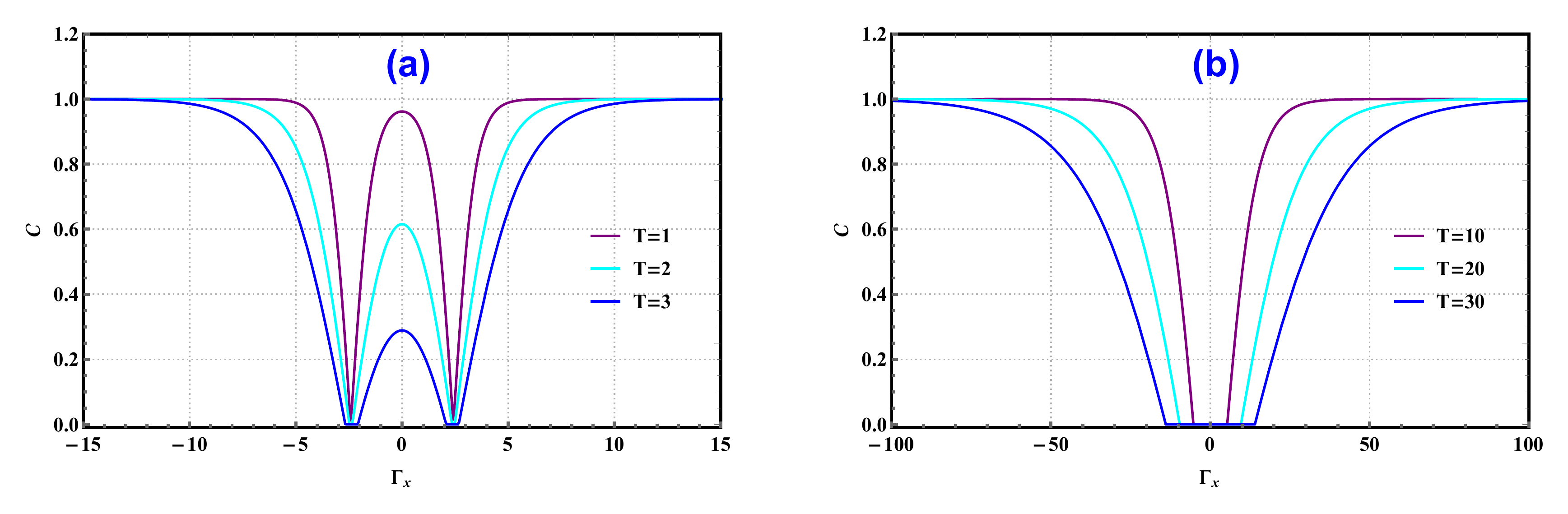}
  \caption{(Color online) The concurrence versus x-component of KSEA interaction for different $T$  ($J=D_x=1$).}\label{f3}
\end{figure}
In Fig. \ref{f3}, we display the concurrence $\mathcal{C}$ in terms of the x-component of the KSEA interaction $\Gamma_x$ at low temperature for $T=$ 1, 2 and 3 (a) and high temperature for $T=$ 10, 20, and 30 (b). In Fig. \ref{f3}.b, the first finding is that for $\Gamma_ x=0$, the concurrence is symmetrical. For the large absolute value of the x-component of the KSEA interaction at the low temperature shown in Fig. \ref{f3}.a, the concurrence tends to a constant value of $\mathcal{C}=1$ even as the temperature increases, i.e., the system's ground state for $\Gamma _x >0 $ is a maximally entangled state $\varphi_2$ (the yellow area in Fig. \ref{diag}.a), whereas for $\Gamma _x <0 $, the ground state is $\varphi_1$ (the green area in Fig. \ref{diag}.a), which is likewise a maximally entangled state, implying that $\mathcal{C}=1 $. Furthermore, for $2<\Gamma _x <3$ ($-3<\Gamma _x <-2$), the ground state of the system is an equal mixture of the doublet states $\varphi_2$ and $\varphi_4$ ($\varphi_1$ and $\varphi_4$), seen in the lines delimiting distinct ground states in Fig. \ref{diag}.a, thus involving to $\mathcal{C}=0$. Additionally, we see that entanglement rises symmetrically, and thus with a drop in temperature, we may encounter a brief temporal entanglement. Otherwise, the concurrence displays an interval in which $ \mathcal{C} $ remains zero at high temperatures near $\Gamma _x = 0 $, and the width of this interval is dependent on the temperature. Thus a temperature rise suggests a lengthening of this interval. In conclusion, the KSEA interaction influences the state's system. Large values of $\Gamma _x$ make the system more entangled, while small values of $\Gamma _x$ make the system less entangled at high temperatures, implying that the system's states become separable. Furthermore, at high temperatures, the DM and KSEA interactions have the same impact on the concurrence ( Fig. \ref{f2}.b and Fig. \ref{f3}.b.)

\section{Conclusion and perspectives}
{\color{red}Concurrence, a measure of entanglement, is investigated in a two-qubit Heisenberg XXX chain with x-components of DM and KSEA interactions.} The Hamiltonian model is given, and through mathematical calculations, the eigenstates entanglement has been determined, and the thermal state at a finite temperature is explicitly derived. The entanglement of the ground state at zero temperature limits and the associated phase diagram has been discussed. Afterwards, we obtained the concurrence expression, which depends on the spin's coupling exchange constant $J$, the DM and KSEA interactions' x-components, and the temperature $T$. Subsequently, we have studied various limits.
Indeed, we analyzed the case at high temperature, the strong and weak spin coupling exchange J compared to the couplings of DM and KSEA. The behaviors of the concurrence-measured entanglements for our study have been investigated numerically. In this paper, we have concluded that temperature $T$, the
x-components of the DM and KSEA interactions, may all play a role in determining the degree of intricacy between the states to a greater or lesser extent. Moreover, it is possible to infer from these results that the separability of the states can be obtained in the high-temperature domains, $J = 0$, or at high temperature for small values of the constants of DM or KSEA interactions. Again, the entanglement of the states can be obtained for large values x-components of the DM and KSEA interactions or low temperature. Moreover, DM and KSEA interactions similarly affect concurrence behaviors at high temperatures.

Still some interesting questions to be addressed. Can we use the studied system to investigate entanglement's dynamic behaviors and describe the basic features of the quantum entanglement at a finite time \cite{Gonzalez}? A related question arose, what about other correlation measurements to study? These issues and associated questions are under consideration.
\section*{Acknowledgment}

Our sincere gratitude goes to Mohamed Monkad, head of the Laboratory for Physics of Condensed Matter (LPMC) in the Faculty of Sciences at Choua\"ib Doukkali University, for his essential assistance.
\\

\textbf{Funding:} Not applicable.
\\

\textbf{Data availability:} Not applicable.
\\

\textbf{Code availability:} Not applicable.
\section*{Declarations}

\textbf{Conflict of interest:} The authors declare that they have no conflict of interest.

\end{document}